
\pageno=0
\magnification=\magstep1
\baselineskip=13pt
\vsize=8.5truein
\hsize=6.5truein
\parskip=7pt
\rightline{McGill/91-34}
\medskip
\rightline{January 1992}
\bigskip
\bigskip
\centerline{\bf CAN A LATTICE STRING HAVE A VANISHING}
\medskip
\centerline{\bf COSMOLOGICAL CONSTANT?}
\bigskip
\bigskip
\centerline{Terry Gannon}
\centerline{Mathematics Department, Carleton University}
\centerline{Ottawa, Ont., Canada  K1S 5B6}
\medskip
\centerline{{\it and}}
\medskip
\centerline{C.S. Lam}
\centerline{Physics Department, McGill University}
\centerline{Montreal, P.Q., Canada~~H3A 2T8}
\bigskip\bigskip
\vskip 3truecm
\noindent
\centerline{\underbar{\bf Abstract}}
\bigskip
We prove that a class of one-loop partition functions found by
Dienes,
giving rise to a vanishing
cosmological constant to one-loop, cannot be realized by a consistent
lattice string.
The construction of non-supersymmetric string with a vanishing
cosmological constant
therefore remains as elusive as ever. We also discuss a new  test
that any
one-loop
partition function for a lattice string must satisfy.

\vfill\eject

\overfullrule=0pt

\def\ie{{\it i.e.}$\,\,\,$}
\def\eg{{\it e.g.}$\,\,\,$}
\def\sp{\,\,\,}
\def\R{{\bf R}}
\def\sp{\,\,\,}
\def\l{\Lambda}
\def\.{\cdot}
\def\Z{{\bf Z}}

\def\equi{\,{\buildrel \rm def \over =}\,}

\def\t{\tau} \def\tb{{\bar \tau}}
\def\.{\cdot}

\def\h{\theta}
\def\T{\Theta} \def\u{\tau} 
\def\b#1{\bar #1}
\def\bt{\bar\tau}

\centerline{{\bf 1. Introduction}}  \bigskip

One of the most serious problems of the current theory of particle
physics
is its inability to account in a natural way for the large size of
the universe and the smallness of the cosmological constant
$\Lambda_{cos}$.
If we ignore the condensates, then the cosmological constant is
computed from
vacuum loops.
 In ordinary quantum field theory, vacuum loops are infinite
so this computation  cannot be carried out reliably. Even if an
ultraviolet
cutoff is put
in to render them finite, a fine-tuning of the cutoff to many many
orders of
magnitude will be needed to obtain the observed cosmological
constant.
Baby universes [1] have been suggested as a mechanism to give a small
cosmological constant, but we will not discuss that scenario here.

Since fermion loops give an opposite sign than boson loops, vacuum
loops
can be rendered small, or even zero, by a suitable combination of
fermionic and bosonic contributions. It is this classical mechanism
to one-loop order which we will discuss in this paper.
The combined contribution is zero for
a supersymmetric theory,
 but unfortunately our world is
not supersymmetric
--- not to the required accuracy  of millivolts anyway. Nevertheless,
there are
still
infinitely many other ways to arrange such a cancellation between the
fermionic
and the bosonic contributions.

In a superstring, the vacuum loops are finite. Moreover,
the bosonic and fermionic mass spectra are highly constrained,
so
it becomes possible and interesting to ask whether a
non-supersymmetric string
theory can give rise to a zero cosmological constant.
The cosmological constant in a string theory can be computed by
integrating (over the modular-parameter $\t$ in the fundamental
region of the modular group) the
partition function of the string, so the problem becomes that of
finding the
right partition functions with a zero integral.

There have been various approaches towards resolving the cosmological
constant problem within the context of string theory.  Taylor and
Itoyama [2] found that $\Lambda_{cos}$ is exponentially suppressed
if the string spectrum has equal numbers of massless bosons
and fermions;  however, the vast majority of self-consistent string
models do not have this property.
Ginsparg and Vafa [3] examined the case of toroidally compactified
strings, and found that $\Lambda_{cos}$ is extremized when these
compactified theories have enhanced gauge symmetries ({\it i.e.,} at
special background-field expectation values).
Unfortunately, no cases were found for which the extremized value
was zero.
A more interesting suggestion was that of Moore [4], who
proposed a certain modular-form symmetry known as Atkin-Lehner
symmetry
as a mechanism for obtaining
a vanishing cosmological constant in the absence of spacetime
supersymmetry.
The basic idea is that any string theory giving rise to an
Atkin-Lehner-symmetric partition function
would have a vanishing one-loop cosmological constant.
Unfortunately,
Moore [5], Taylor [6], and Schellekens [7] were unable
to construct string models in
$D>2$ spacetime dimensions which give rise to such partition
functions, and
Balog and Tuite [8] succeeded in proving that
no such string models can exist.  Dienes [9] then
was able to generalize the Atkin-Lehner mechanism
in a variety of ways such that the conclusions of
the Balog-Tuite no-go theorem are avoided.
It therefore remains an open question as to whether string models
exist with partition functions displaying these generalized
Atkin-Lehner
symmetries.

Another promising approach was obtained by Dienes [10] recently,
who  found a class of modular-invariant partition functions, given in
eqs.~(1$a,b$) below,
 which gives rise to a  zero
cosmological constant to one-loop order, though this class of
partition
functions
does not exhibit an Atkin-Lehner or generalized Atkin-Lehner
symmetry.
Moreover, his partition
functions
satisfy a number of additional constraints (\eg they have no on-shell
tachyons) which physically acceptable strings are expected to obey.
The partition functions he found are the kind that one would obtain
from a
lattice string [11], but after looking over more than
120 000 such strings with the help of a computer,
 he
was  unable to find a consistent string with such a partition
function [10]. We will show in this paper that a string within this
class of
partition
functions does not exist. As a result, the construction of a
non-supersymmetric
string
with a zero cosmological constant remains as elusive as ever.
Dienes' proposal  is briefly summarized in Sec.~2 below.

A lattice string [11,15] is defined by a self-dual
lattice $\l$ of signature (22,10), and a fermionic vector $v$, and
its
partition function is modular invariant. On the other hand, modular
invariant partition functions such as Dienes' do not have to come
from a single lattice $\l$, since for instance linear combinations
 of modular-invariant functions are modular-invariant.
In this paper we
apply lattice techniques  to investigate the
question whether any consistent lattice string can be found
corresponding
to one of Dienes'
partition functions.
 We will be able to quickly rule out in Sec.~3 half of his class.
 In Sec.~4 we show that strings cannot be constructed from
  the remaining half of his class either,
 provided that the
strings satisfy the {\it half-norm property} (given in eq.~(9)).
This
property is extremely natural given the class of possible partition
functions,
and indeed is consistent with the type of string with which Dienes
was concerned. In Ref.~[12] we shall consider what can be proved when
we drop
that
restriction and consider instead all conceivable strings.

As a result of this study, we have also come up with a new test any
partition
function
coming from a lattice string must satisfy. This is contained in the
Corollary
in Sec.~4.
This test can be used to check whether an interesting looking
partition
function could come
from a lattice string.

Some of the terminology of lattices used here can be found in the
Appendix.
For an introduction to the basic theory of lattices, see \eg Ref.~13
and
particularly Ref.~14. A discussion of the lattice string can be found
in Ref.~15.

\bigskip \bigskip \centerline{{\bf 2. Dienes' class of partition
functions}}
\bigskip

Dienes'  one-loop partition functions are ${\rm Im}(\tau)^{-1}
\eta(\tau)^{-24}\overline{\eta(\tau)^{-12}}T(\tau,\b\tau)$,
for $T$ given by
$$T(\tau,\b\tau)=cQ(\tau,\b\tau)+I(\tau,\b\tau),\eqno(1a)$$
where $Q$ is given by $$\eqalignno{
Q(\tau,\b\tau)&=\b\h_2^2\h_2^2\{\h_2^4\h_3^4\h_4^4[2\h_3^4\h_4^4
\b\h_3^4\b\h_4^4
-\h_3^8\b\h_4^8-\b\h_3^8\h_4^8]+\h_2^{12}[4\h_2^8\b\h_3^4\b\h_4^4+13
\h_3^4\h_4^4\b\h_3^4\b\h_4^4]\}&\cr
&+\b\h_3^2\h_3^2\{\h_2^4\h_3^4\h_4^4[2\h_2^4\h_4^4\b\h_2^4\b\h_4^4
-\h_2^8\b\h_4^8-\b\h_2^8\h_4^8]+\h_3^{12}[4\h_3^8\b\h_2^4\b\h_4^4-13
\h_2^4\h_4^4\b\h_2^4\b\h_4^4]\}&\cr
&+\b\h_4^2\h_4^2\{\h_2^4\h_3^4\h_4^4[2\h_2^4\h_3^4\b\h_2^4\b\h_3^4
-\h_2^8\b\h_3^8-\b\h_2^8\h_3^8]+\h_4^{12}[4\h_4^8\b\h_2^4\b\h_3^4+13
\h_2^4\h_3^4\b\h_2^4\b\h_3^4]\},&\cr
&&(1b)
}$$
and where
$I$ is any arbitrary modular-invariant  function of $\t$ and $\tb$
with the
property that
the Taylor expansion $\sum_{m,n} a_{mn}{\bar q}^m q^n$ of
$\eta(\t)^{-24}
\overline{\eta(\t)^{-12}}I(\tau,\b\tau)$ satisfies $a_{mn}=-a_{nm}$.
Here and throughout
this paper, $q\equi \exp(\pi i \tau)$ and
 $\b q\equi\exp(-\pi i \tb)$.
The functions $\h_i\equi \h_i(\t)$ and $\bar\h_i\equi
\h_i(-\tb)=\overline{\h_i(\t)}$  are
Jacobi theta functions defined in eqs.~($A.1$), and $\eta$ denotes
the Dedekind
eta function.
  Hence $I$ is of the form $(\h_2\h_3\h_4)^4[X(\t,\tb)-X(-\tb,-\t)]$.
Functions $I$ of this type will not contribute to the cosmological
constant, and hence can be left arbitrary, but they certainly could
affect whether a lattice corresponding to the partition function
$T(\tau,\b\tau)$ of ($1a$) could be found.

Note that eq.~(1$b$) implies the string has 22
left-moving bosonic degrees of freedom and 10 right-moving ones.

The constant $c$ in $(1a)$ is limited by the off-shell tachyon
constraint of
Ref.~[10] to be a rational number of the form $n/32$, with $1\le n\le
10$.
We can derive this [12,16] simply by counting unit vectors without
using the off-shell tachyon constraint, but this restriction on $c$
is not
needed
in this paper save for a convenient classification of Dienes'
partition
functions
into two classes which we will consider in Secs.~3 and 4
respectively.
In Sec.~3, we shall show that a lattice string cannot be constructed
if
$|c|> 5/32$. In Sec.~4, we will show that a lattice string cannot be
constructed for any $c$ if the {\it half norm property} (discussed in
eq.~(9)
below) is satisfied.

\bigskip \bigskip \centerline{{\bf 3.  The $|c|>5/32$ case}}
\bigskip

 Before proceeding to the proof, let us first define a few things to
establish
notation. Also see the Appendix for further details.

A lattice $\l$ shall be called $v$-{\it even} for some vector $v$
(not
necessarily in $\l$) if
$$r^2+2r\cdot v\equiv 0 \sp ({\rm mod}\sp 2) \sp\forall
r\in\l.\eqno(2)$$
It is easy to show that
  a lattice $\l$ can be  $v$-even for some $v$ only if $\l$ is
integral;
whenever  $\l$ is integral, such a  $v$ can always be found, and must
lie in
${1\over 2}\l^*$.

For any Euclidean lattice $\l$ define the {\it shifted theta
constant}
$\Theta_{\l}(vu|\t)$ to be
$$\eqalignno{\Theta_{\l}(vu|\t)&=\sum_{r\in \l}\exp[\pi i \t
(r+u)^2+2\pi i
(r+u)\cdot v].& (3a)}$$
 For an indefinite lattice $\l$, the {\it shifted theta constant} is
defined by
$$\Theta_{\l}(vu|\tau,\b\tau)=\sum_{r\in \l}\exp[\pi i \t
(r_L+u_L)^2-\pi i
\tb (r_R+u_R)^2+2\pi i (r+u)\cdot v], \eqno(3b)$$
where we write $r\in\l$ in the usual way as $r=(r_L;r_R)$ (so dot
products in
$\l$ are given by $r\cdot r'=r_L\cdot r'_L-r_R\cdot r'_R$). We will
also use
the short-hand $\Theta(\l)(\tau,\b\tau)$ for
$\Theta_{\l}(00|\tau,\b\tau)$ and
$\Theta({\l})(\t)$ for $\Theta_\l(00|\tau)$.
They
will sometimes be called {\it pure theta constants}.

We know  that if Dienes' partition function corresponds to a
consistent lattice string [11,15], then we must have
$$T(\tau,\b\tau)=\Theta_{\l}(vv|\tau,\b\tau),\eqno(4)$$ for some  odd
indefinite
$v$-even self-dual lattice $\l$ of signature (22,10), and some {\it
fermionic vector} $v=(0;v_R)$ satisfying $v^2=-v_R^2=-1$. The
fermionic vector is there to provide fermionic quantum numbers for
the string; it is given by the
 projection of an internal momentum  along the fermionic vector.

Given any indefinite lattice $\l$ of signature $(m,n)$, define the
$n$-dimensional  Euclidean lattice $\l_R$, called the {\it
RHS} of $\l$, by:
$$\eqalignno{\l_R&\equi \{r_R|(0;r_R)\in \l\}, &(5)\cr }$$
where the Euclidean dot products in $\l_R$ are given  by $r_R\cdot
r'_R=-(0;r_R)\cdot (0;r'_R)$. Define similarly  the $m$-dimensional
Euclidean
lattice $\l_L$ (the {\it LHS} of $\l$).

We would like to find the lattice $\l$ responsible  (in the sense of
eq.~(4))
for
the partition function eq.~(1), or show that no  such lattice exists.
One
glaring
difficulty is that the partition function is not precisely known
because of the arbitrary nature of $I(\tau,\b\tau)$ in (1). We will
now
discuss two methods of overcoming that difficulty.

 The first method involves looking at the shifted theta constant for
$\l_R$,
which can be obtained from the shifted theta constant for $\l$ by
 putting $q$ equal to zero (\ie considering the limit
 $\t \rightarrow +\infty i)$ in  eq.~$(1a)$, because any acceptable
$I$ in
eq.~(1) vanishes in this limit.
 From $(1a)$ and (4),  and using the Jacobi identity, we obtain
 $$\eqalignno{\Theta_{\l_R}(v_Rv_R|\tb)=&4c[\b\theta_3^8\b\theta_4^2+
 \b\theta_3^6\b\theta_4^4-\b\theta_3^4\b\theta_4^6-
\b\theta_3^2\b\theta_4^8]&\cr
 =&16c\{8{ \b q}-896{ \b q}^5+5184{\b q}^9+\cdots\}.&(6)\cr}$$

Next, consider the symmetrization
$$\tilde{\Theta}_{\l}(vv|\tau,\b\tau) \equi
\h_2^4(-\tb)\h_3^4(-\tb)\h_4^4(-\tb)\Theta_{\l}(vv|\tau,\b\tau)
+\h_2^4(\t)\h_3^4(\t)\h_4^4(\t)\Theta_{\l}(vv|-\tb,-\t).\eqno(7a)$$
By definition the $I$ in eq.~(1) satisfy $\tilde{I}=0$, so we can
similarly
write down explicitly an expression for
$\tilde{\Theta}_{\l}(vv|\t\bt)$ in
 terms of the Jacobi functions (we use the Jacobi identity in the
Appendix  to
rewrite this polynomial so that it is
of degree $\leq 2$ in both $\h_2$ and $\b\h_2$. See also the remark
in the
second paragraph of Sec.~4). The result is very
messy and is given in Ref.~[16], but fortunately we only need two of
the
terms:
$$\tilde{\Theta}_{\l}(vv|\t\bt)=-13c\,\h_3^{18}\h_4^4\b\h_3^{14}\b\h_
4^8
-11c\,\h_3^{18}\h_4^4\b\h_3^6\b\h_4^{16}+\cdots.\eqno(7b)$$

In particular, note that the coefficients in eq.~(7$b$) of $\h_3^{18}
\h_4^4\b\h_3^{14}\b\h_4^8$ and $\h_3^{18}\h_4^4\b\h_3^6\b\h_4^{16}$
are not
equal. We will show in Sec.4 that those coefficients must be equal
for any lattice satisfying the half-norm property eq.~(9).

 Rather than directly trying to solve eqs.~(1) and (4) for $\l$, it
will be
simpler first to try to solve eqs.~(6)
 and (7$b$) for $v_R$, $\l_R$, and ultimately $\l$ (or show that no
such
solution exists).

One obvious difficulty one encounters is that eqs.~(6) and ($7b$)
involve
`shifted theta
constants' $\Theta_{\l_R}(v_Rv_R|\t)$ {\it etc.}, rather than {\it
pure} theta
 constants. This makes it much harder to read off information about
 $\l_R$ and $\l$ because the fermionic vector $v$ or $v_R$ is not
{\it a
priori}
known. Eqs.~(8) below are designed to overcome this
complication. With this in mind, make the following definitions.

Let $\l_B$ be the sublattice of $\l$ consisting only of the even norm
vectors
(\ie the {\it bosons}).  Then it can be shown that its determinant is
$|\l_B|=4$, and its
 signature is still (22,10). Let $\l_{BR}\equi(\l_B)_R$.
Let $u=(0;u_R)\in \l$
be any odd-normed vector living entirely in the right-hand side of
$\l$ (such
vectors will always exist, by eq.~(6)).  Then $\l=\l_B[u]$ and
$\l_R=\l_{BR}[u_R]$ (see the Appendix for this gluing notation). Note
also that $2v,2u\in \l_B$ and
$2v_R,2u_R\in \l_{BR}$.

Now define lattices $\l'_R\equi \l_{BR}[v_R+u_R]$ and $\l''_R\equi
\l_{BR}[v_R]$; define $\l'$ and $\l''$ similarly.
 Note that
$\l_R$, $\l'_R$ and $\l''_R$ are all integral (in fact, odd) and have
equal
determinant.    Both
$\l_R$ and $\l_R'$ are $v_R$-even, and $\l_R''$ is
$u_R$-even.

An explicit calculation, using eqs.~(3) and some formulas
in Ref.~[10], yields:

 $$\eqalignno{
\Theta_{\l_R}(v_Rv_R|\tb)&=\Theta({\l_R''})(\tb)-\Theta({\l_R'})(\tb)
,
&(8a)\cr
\Theta_{\l}(vv|\tau,\b\tau)&=\Theta({\l''})(\tau,\b\tau)-\Theta({\l'}
)(\tau,\b\tau).
&(8b)\cr}$$
 Note that eqs.~(6) and ($8a$) imply that the number of norm 1
vectors in
$\l_R''$
minus the number in $\l_R'$ must equal $128c$. However it is easy to
see that
$\l_R''$ and $\l_R'$, being integral, Euclidean and 10-dimensional,
can both
have at most 20 norm 1 vectors. Therefore $-20\leq 128c\leq 20$, \ie
$0\neq
|c|\leq 5/32$.

This immediately gives us the first main result of this paper:

\bigskip \noindent{\bf Theorem A}: \quad No lattice string exists
having a
 partition function $T$ in Dienes' class  (eq.~(1)) if $|c|>5/32$.
  \bigskip

\bigskip \bigskip \centerline{{\bf 4. The half-norm case}} \bigskip

We shall now proceed to consider the case of a general $c$. One way
is to
enumerate
all solutions $\l_R,v_R,c$ to eq.~(6), and then proceed to study
whether they
can satisfy
(1). The enumeration can be done if (the determinant) $|\l_R|<64$ ---
there are
exactly three of these up to
integral
equivalence, and they are listed in Ref.~[16].
However, a complete enumeration of all $\l_R$ satisfying eq.~$(6)$
could be
unfeasible. Nevertheless a large class of solutions $\l_R$
to eq.~(6), which includes all such solutions we have ever found,
satisfies the following property, which we shall call
the {\it half-norm property}:
$$g\in \l_R^* \Rightarrow g^2\in {1\over 2}\Z.\eqno(9)$$
Indeed, it may turn out that {\it any} solution
of eq.~(6) must satisfy this additional property, since the
contributions
to expressions such as eq.~($7b$) of glue vectors $(g_L;g_R)\in \l$
with $g_R$ violating eq.~(9) (\ie with $g_R^2 \notin {1\over 2}\Z$)
would otherwise have to conveniently cancel out.
In any case, strings with the spin structures
considered
in Ref.~[10] all seem to satisfy this property. We shall therefore
assume the half-norm property to be valid from now on; this
assumption
is dropped in Ref.~[12].

Now we need to use a mathematical theorem:
Cor.~10.2 in Ref.~[18]. Together with the half-norm assumption,
 this corollary implies that
 the theta constants of all glue classes in
$\l_R'{}^*/\l_R'$ and $\l_R''{}^*/\l_R''$ can be expressed as
polynomials in $\h_2^2, \h_3^2, \h_4^2$,
and that there is only one such
form for each such polynomial subject to the additional condition
that
it be of degree $\leq 2$ in $\h_2$. It can be shown, using for
example
Thm.~2.5 of Ref.~[19], that any integral lattice $\l$ satisfying
eq.~(9)
also has the property that any glue $g\in\l^*$ must be of order 1,2
or 4. This immediately implies that the determinant $|\l|$ must be
a power of 2; if
it is to also be a solution of eq.~(6) it can be shown that the
determinant
must
be a power of 4. Also note that if one of $\l_R$, $\l_R'$, $\l_R''$
satisfies
eq.~(9), all do (see eq.~(10)).

There are several simultaneous solutions [16] $\l_R$ to eqs.~(6) and
(9).
Their
determinants range from $4^2=16$ (for which there are 3 solutions) to
$4^8=$16
384 (with 4 solutions). It is unnecessary to explicitly find any of
these,
however. In the following paragraphs we will show that the partition
function
eq.~(4) of any self-dual $v$-even $\l$
of signature (22,10) whose RHS $\l_R$ satisfies eq.~(9), cannot
be in the class eq.~(1$c$). In particular, we will show that in the
expansion
of $\tilde{\Theta}_{\l}(vv|\t\bt)$, the coefficients of
$\h_3^{18}\h_4^4
\b\h_3^{14}\b\h_4^8$ and $\h_3^{18}\h_4^4\b\h_3^6\b\h_4^{16}$ must be
equal,
thus violating eq.~(7$b$).

The motivation for this approach was obtained by explicitly computing
the
partition functions corresponding to some explicit solutions $\l$ to
eq.~(6),
and comparing with eq.~(7$b$). An example of such a calculation was
included
in Ref.~[16].

Let $\l$ be any self-dual, $v$-even lattice of signature (22,10)
whose RHS
$\l_R$ satisfies eq.~(9). Let $D=|\l_R|$. Enumerate the $D$ glue
classes
$[g_i]\l_R$ of $\l_R^*/\l_R$. Without loss of generality (by
replacing
$g_i$ if necessary with $g_i+u_R$) we may
choose the
representatives so that $g_i\cdot v_R \in \Z$. Define
$g_i'=g_i+v_R+u_R$,
$g_i''=g_i+v_R$; then $[g_i']\l_R'$ and $[g_i'']\l_R''$ for
$i=1,\ldots,D$
exhaust the $D$ glue classes of $\l_R'{}^*/\l_R'$ and
$\l_R''{}^*/\l_R''$.
Note that the correspondences $g_i \leftrightarrow g_i'
\leftrightarrow g_i''$
define group isomorphisms $\l_R^*/\l_R\cong \l_R'{}^*/\l_R' \cong
\l_R''{}^*/\l_R''$.

Now consider the LHS $\l_L$ of $\l$ --- it too has $D$ glue classes
in $\l_L^*/\l_L$, by \eg Thm.2.4 of Ref.~[15]. They can be enumerated
in such a way that $$\l=\bigcup_{i=1}^D \bigl([h_i]\l_L;
[g_i]\l_{R}\bigr).$$ Then $h_i \leftrightarrow g_i$ allows us to
extend the
above glue group isomorphisms. We also get the following congruences:

$$h_i \cdot h_j \equiv g_i \cdot g_j \equiv g_i'\cdot g_j' \equiv
g_i'' \cdot
g_j'' \sp ({\rm mod}\sp 1),\sp 1\leq i,j\leq D.\eqno(10)$$

A simple calculation establishes the gluings
$$\l'=\bigcup_{i=1}^D \bigl([h_i]\l_L;
[g_i']\l_{R}'\bigr) \sp {\rm and} \sp \l''=\bigcup_{i=1}^D
\bigl([h_i]\l_L;
[g_i'']\l_R''\bigr).$$
These immediately imply
$$\Theta_{\l'}(\t\bt)=\sum_{i=1}^D \Theta([h_i]\l_L)(\t)\cdot
\Theta([g_i']\l'_{R})(\b\t),\eqno(11a)$$
with a similar expression for $\Theta_{\l''}$. Hence
$$\eqalignno{\Theta_{\l}(vv|\t\bt)=&\sum_{i=1}^D
\Theta([h_i]\l_L)(\t) \cdot
\{\Theta([g_i'']\l_R'')(\bt)-\Theta([g_i']\l_R')(\bt)\}& \cr
\equi & \sum_{i=1}^D \Theta([h_i]\l_L)(\t)\cdot
\Delta_i(\bt),&\cr&&(11b)
\cr
\tilde{\Theta}_{\l}(vv|\t\bt)=& \sum_{i=1}^D \{\Theta([h_i]\l_L)(\t)
\cdot
\Delta_i(\bt) \cdot \b\h_2^4 \b\h_3^4
\b\h_4^4+\Theta([h_i]\l_L)(\bt)\cdot
\Delta_i(\t) \cdot \h_2^4 \h_3^4 \h_4^4 \}. &(11c) \cr}$$

As mentioned above, Cor.10.2 in Ref.~[18] tells us we can
{\it uniquely} write eq.~(11$c$) as a polynomial in the Jacobi
functions so
that it is of degree $\leq 2$ in both $\h_2$ and $\b\h_2$. We are
trying to
show that the coefficients of $\h_3^{18}\h_4^4\b\h_3^{14}\b\h_4^8$
and
$\h_3^{18}\h_4^4\b\h_3^6\b\h_4^{16}$ in eq.~(11$c$) --- call them $A$
and $B$
--- are equal.

 Note that $A=A_1-A_2+A_3$, where $A_1,A_2,A_3$ are, respectively,
the
coefficients of $\h_3^{18}\h_4^4\b\h_3^{6}\b\h_4^4$,
$\h_3^{18}\h_4^4\b\h_3^{10}$ and $\h_3^{14}\h_4^8\b\h_3^{10}$ in
$\Theta_{\l}(vv|\tau,\b\tau)$; similarly, $B=-B_1+B_2$ where
$B_1,B_2$ are,
respectively, the coefficients of $\h_3^{18}\h_4^4\b\h_3^{2}\b\h_4^8$
and
$\h_3^{6}\h_4^{16}\b\h_3^{10}$ in $\Theta_{\l}(vv|\tau,\b\tau)$.

Now, consider any
$\Delta_k=\Theta([g_k'']\l''_{R})-\Theta([g_k']\l'_{R})$ for
which $g_k^2\in {\bf Z}$. Then each $\Delta_k$ can be expressed as a
polynomial
in $\b\h_3^2$ and $\b\h_4^2$. Let $\Delta_k'$ consist of those terms
in $\Delta_k$ in which $\b\h_3^2$ occurs to odd power. For example,
$(\Theta_{\l_R})'=4c\b\h_3^6 \b\h_4^4-4c\b\h_3^2
\b\h_4^8$, by eq.~(6). {\it A priori} one would expect these
$\Delta_k'$ to look
like $r_k\b\h_3^{10}+s_k\b\h_3^6\b\h_4^4+t_k\b\h_3^2\b\h_4^8$, for
arbitrary
$r_k,s_k,t_k \in {\bf R}$. However, if we can show that for each of
these $k$
there exists an $\ell_k\in \R$ such that
$$\Delta_k'(\bt)=\ell_k(\b\h_3^6
\b\h_4^4-\b\h_3^2 \b\h_4^8), \eqno(12)$$ then it is easy to see that
 $A_1=-B_1$ and $A_2=A_3=B_2=0$, \ie that $A=B$. Hence it suffices to
show that
eq.~(12) holds for each $k$ for which $g_k^2\in \Z$.

This leads us to the second main result of this paper.

\bigskip
\noindent{\bf Theorem B}:\quad There is no string theory with
partition
function of the type given in eq.~(1$a$), based on a lattice $\l$
whose RHS
$\l_R$ satisfies the half-norm property, \ie eq.~(9). \bigskip

\noindent{\it Proof} \quad
We will begin by making some general observations about the theta
constants of
lattices satisfying eq.~(9). Only in the final paragraph of the proof
will it be
applied to $\l_R'$ and $\l_R''$.

Let $\l_1$ be any 10-dimensional (integral) lattice satisfying
eq.~(9).
Let $D= |\l_1|$. We can write $$\Theta(\l_1)=
a\h_3^{10}+b\h_3^8\h_4^2+c\h_3^6\h_4^4+d\h_3^4\h_4^6+e\h_3^2
\h_4^8+f\h_4^{10},\eqno(13a)$$ where $a,b,c,d,e,f$
are real, and $f=1-a-b-c-d-e$. Then $$\eqalignno{
\Theta(\l_1^*)=&
Da\h_3^{10}+Db\h_3^8\h_2^2+Dc\h_3^6(\h_3^4-\h_4^4)&\cr
&+Dd\h_3^4\h_2^6
+De\h_3^2(\h_3^4-\h_4^4)^2+Df\h_2^{10}&(13b)
\cr =&(Da+Dc+De)\h_3^{10}+Db\h_3^8\h_2^2
+(-Dc-2De)\h_3^6
\h_4^4&\cr &+Dd\h_3^4\h_2^6+De\h_3^2\h_4^8+Df\h_2^{10}&(13c)}$$ (see
Ref.~[17]).
Because $\l_1^*$ only has one zero
vector, eq.~(13$b$) implies $a=1/D$.

Now let $g\in \l_1^*$, $g^2\in {\bf Z}$.
Then $g$ will be order 1,2 or 4, and $\Theta([g]\l_1)$ will be of the
same form as eq.~(13$a$): \ie
$\Theta([g]\l_1)=a_g\h_3^{10}+b_g\h_3^8\h_4^2+\cdots$.
 Of course, $\l_1[g]$
will also satisfy eq.~(9). Consider first the case where $g$ is of
order 2. Then
the previous paragraph applied to both $\l_1$ and $\l_1[g]$
immediately implies
that $a_g=1/D$. Hence the same conclusion must apply to $g$ of order
4 (and
trivially to order 1 glues) --- \ie the leading coefficient for any
integral-normed
glue $g$ of $\l_1$
is $a_g=1/D$. Moreover, note that the coefficient of  $\h_3^{10}$ in
the theta constant of a glue class of
non-integral norm must be 0.

Of course,
we can rewrite eq.~(13$c$) as the sum of $\Theta([g]\l_1)$ for all
glue
classes $[g]\l_1\in \l_1^*/\l_1$. We then get $N/D$ as the
coefficient of
$\h_3^{10}$ there, where $N$ is the number of integral-normed glue
classes
of $\l_1$.
Hence $c+e=(N-D)/D^2$. The same technique as in the previous
paragraph allows
us to find a similar formula for $c_g+e_g$, for glues $g$ of $\l_1$
with integral norm.

Finally, consider the two lattices $\l_R'$ and $\l_R''$ corresponding
to an integral
$\l_R$ satisfying eq.~(9). Their glues $g_i',g_j''$ can be paired as
was done
around eq.~(10).
Consider  integral-normed glues $g_k'\leftrightarrow
g_k''$. From the previous two paragraphs (and the dot
product-preserving glue
group isomorphism analogous to that defined around eq.~(10)),
two things should be clear: $a_{g_k'}=
a_{g_k''}$ and $c_{g_k'}+e_{g_k'}=c_{g_k''}+e_{g_k''}$. Hence
$$\Delta_k'=(a_{g_k''}-
a_{g_k'})\b\h_3^{10}+(c_{g_k''}-c_{g_k'})\b\h_3^6\b\h_4^4+(e_{g_k''}-
e_{g_k'})
\b\h_3^2\b\h_4^8$$ necessarily satisfies eq.~(12).
\qquad QED \bigskip

The argument considered in this section can be formulated
into a general test. In particular, let $Q_0(\t,\bt)$ be any given
function, and consider the class of partition functions
$$T_0(\t,\bt)=cQ_0(\t,\bt)+I(\t,\bt),$$
where $I=(\h_2\h_3\h_4)^4[X(\t,\tb)-X(-\tb,-\t)]$ is any skew
symmetric
function as defined before, and $c$ is
any nonzero real constant. Then:

\bigskip \noindent{\bf Corollary}:\quad If such a
partition function $T_0$ is to be realizable by a lattice string
defined
by a lattice $\l$ whose RHS satisfies eq.~(9),  the following
two conditions must be satisfied:

\item{(i)} \quad $\tilde{Q_0}$ can be written as a polynomial in
$\h_i$
and $\b \h_i$; and

\item{(ii)} \quad making this polynomial of degree $\leq 2$ in both
$\h_2$ and $\b \h_2$, the coefficients of $\h_3^{18} \h_4^4
\b\h_3^{14}
\b\h_4^8$ and $\h_3^{18} \h_4^4 \b\h_3^6 \b\h_4^{16}$ must be equal.
\bigskip

The symmetrized function $\tilde Q_0$ is obtained  from $Q_0$ as in
$(7a)$.
These conditions are necessary but not sufficient, though
they are enough to rule out the class in eq.~(1).  As we showed in
Sections 3
and 4, Dienes' choice of $Q_0=Q$ satisfies
condition (i) but not (ii). See Ref.~[10] for other conditions
to be satisfied.

\bigskip \bigskip \centerline{{\bf 5. Summary and Conclusions}}
 \bigskip

We have been concerned with the question whether a non-supersymmetric
lattice
string can yield a zero cosmological constant to one-loop order. The
cosmological constant is given by the integral of a partition
function.
Of those partition functions whose integrals vanish,
we have examined a class written down by
Dienes in  [10] (eq.~(1)). These partition functions are of the form
that
might come from a lattice string.

In Sec.~3, we have ruled out any possibility that such partition
functions do
come from
a lattice string if $|c|>5/32$. In Sec.~4, we ruled out all other $c$
values
with the
technical restriction of the `half norm property' (eq.~(9)).
The half-norm property
characterizes the most natural class of possibilities for the lattice
$\l$ in
eq.~(4), since otherwise eq.~(7$b$) (and the equation for
$\Theta(\l_R''{}^*)-
\Theta(\l_R'{}^*)$ that results by replacing $\bt \rightarrow -1/\bt$
in
eq.~(6)) could hold only if the contributions
to their power series made by the glues $(h_i;g_i)$ (and $g_i$,
resp.)
with norms $h_i^2,g_i^2 \notin {1\over 2}{\bf Z}$, were all to
 conveniently cancel.

 We have also devised a new test (the Corollary in Sec.~4) for a
general class
 of partition functions to check whether they can come from any
lattice string.
It is using
 this test that we ruled out above the class in eq.~(1).  The same
test however
can
 be applied to other partition functions as well.

Clearly, our aim now is to extend this proof beyond the half-norm
assumption.
Unfortunately the analysis then becomes more complicated, and will be
included
in a subsequent paper [12]. The goal of that paper is to determine to
what
extent an acceptable solution to eq.~(6) must satisfy eq.~(9), and
thus cannot
be extended to a solution of eq.~(7$b$).

\bigskip \bigskip \centerline{\bf Acknowledgment} This work is
supported in
part by the Natural
Sciences and Engineering Research Council of Canada and the Qu\'ebec
Department of Education.  We thank Keith
Dienes for useful discussions and suggestions.

\bigskip \bigskip \centerline{{\bf Appendix}} \bigskip

Below we define and discuss the various
theta functions used in the text. We will also state some
relevant definitions and results on lattices.

The Jacobi $\theta$-functions which we need are defined by:
$$\eqalignno{
 \h_2(\tau)& \equi
\sum_{m=-\infty}^\infty
q^{(m+1/2)^2},&(A.1a)\cr
\h_3(\tau)& \equi \sum_{m=-\infty}^\infty q^{m^2},
&(A.1b)\cr \h_4(\tau)& \equi \sum_{m=-\infty}^\infty
(-q)^{m^2}, &(A.1c)\cr}$$
where $q\equi \exp(\pi i \tau)$.

Relations abound between $\h_2$, $\h_3$ and $\h_4$ --- see \eg
Ref.~[17]. In
particular: $$\eqalignno{\h_2(\u)&=\h_3(\u/4)-\h_3(\u), &(A.2a)\cr
\h_4(\u)&=2\h_3(4\u)-\h_3(\u), &(A.2b)\cr
\h_3(\u)^4&=\h_2(\u)^4+\h_4(\u)^4. &(A.2c)\cr}$$

We will find it simpler not to avail ourselves of the linear
relations
eqs.~($A.2a,b$). However, eq.~($A.2c)$ (called the {\it Jacobi
identity}) will
be used. The arguments of the $\h$-functions used in this paper will
always be $\u$,
{\it unscaled}, and so we will often suppress the argument.

The {\it glue classes} of lattices are discussed for example in
Ref.~[14]. In short,
$[g]\l_0 \equi g+\l_0$ is called a glue class of a lattice $\l_0$ if
$g\in {\bf Q}
\otimes \l_0$ ({\bf Q} is the set of all rational numbers). For an
integral lattice $\l$, the group $\l^*/\l$ consists
of $D$ glue classes of $\l$, where $D=|\l|$ is the {\it determinant}
of $\l$
(defined as det($e_i \cdot e_j)$, where $\{e_i\}$ is a set of basis
vectors of $\l$), and $\l^*$ is the dual lattice of $\l$.

Given a glue class $[g]\l_0$ of a Euclidean lattice $\l_0$,
 its {\it theta constant} is defined to be
$$\eqalignno{\T([g]\l_0)(\u)&\equi\sum_{x\in\l_0}\exp[\pi i\u
(g+x)^2] &\cr
\T(\l_0)&\equi\T([0]\l_0),&(A.3)}$$
where as usual Im($\u) >0$. See Ref.~[17] for
properties of these functions.

Theta constants of integral lattices and their glue classes are
examples of
modular forms. A discussion of
general results concerning lattice theta constants as modular forms
can be
found on pp.382-8 of Ref.~[13].
\vfill\eject
\parskip=0pt
\centerline{{\bf Bibliography}} \bigskip

\item{[1]}  S. Coleman, {\it Nuclear Physics} {\bf B310}, 643 (1988).
\item{[2]}
H. Itoyama and T.R. Taylor, {\it Phys. Lett.}  {\bf B186}, 129
(1987).
\item{[3]}
P. Ginsparg and C. Vafa, {\it Nucl. Phys.} {\bf B289},  414  (1987).
\item{[4]}
G. Moore, {\it Nucl. Phys.} {\bf B293},  139 (1987).
\item{[5]}
G. Moore, {\it  Nucl. Phys.}  {\bf B299}, 847  (1988).
\item{[6]}
T.R. Taylor, {\it Nucl. Phys.} {\bf B303},  543 (1988).
\item{[7]}
A.N. Schellekens, in {\it Prospectives
in String Theory}, Proceedings of the Niels Bohr Institute Nordita
Meeting, Copenhagen, Denmark, 1987, edited by P. DiVecchia
and J.L. Petersen (World Scientific, Singapore, 1988).
\item{[8]}
J. Balog and M.P. Tuite, {\it Nucl. Phys.}  {\bf B319},  387 (1989).
\item{[9]}
K.R. Dienes, {\it Phys. Rev.} {\bf D42}, 2004  (1990).
\item{[10]}
 Keith R. Dienes, {\it Phys. Rev. Lett.} {\bf
65}, 1979 (1990).
\item{[11]}
H. Kawai, D.C. Lewellen,  and S.-H. H. Tye, {\it Phys. Rev. Lett.}
{\bf 57},
1832 (1986);
\item{} W. Lerche, D. L\"ust, and A.N. Schellekens, {\it Nucl. Phys.}
{\bf
B287}, 477 (1987);
\item{} I. Antoniadis, C. Bachas, and C. Kounnas, {\it Nucl. Phys.}
{\bf B289},
87 (1987);
\item{} C.S. Lam, {\it Int. J. Mod. Phys.} {\bf A3}, 913 (1988).
\item{[12]} Terry Gannon and C.S. Lam, (in progress).
\item{[13]}  J.W.S. Cassels, {\it Rational Quadratic Forms},
(Academic Press, 1978).
\item{[14]}  J.H. Conway and N.J.A. Sloane,  {\it Sphere Packings,
Lattices
and Groups}, (Springer-Verlag, 1988).
\item{[15]}  Terry Gannon and C.S. Lam, {\it Phys. Rev.} {\bf D41},
492 (1990).
\item{[16]}Terry Gannon, {\it Lattices and Theta Functions}, (Ph.D.
Thesis, McGill University, 1991).
\item{[17]}  Terry Gannon and C.S. Lam, {\it Lattices and
$\Theta$-function Identities I:  Theta Constants}, McGill University
preprint No. McGill/91-10, to be published in {\it J. Math. Phys.};
\item{ }  Terry Gannon and C.S. Lam, {\it Lattices and
$\Theta$-function Identities II:  Theta Series}, McGill University
preprint No. McGill/91-11, to be published in {\it J. Math. Phys}.
\item{[18]}  D. Mumford, {\it Tata Lectures on Theta}, Vol. I,
(Birkhauer, 1984).
\item{[19]} Terry Gannon and C.S. Lam, {\it Rev. Math. Phys.} {\bf
3},
331 (1991).

\end